\documentclass[twocolumn,prl]{revtex4}
\usepackage{amsmath}
\usepackage{amssymb}
\usepackage{stmaryrd}
\usepackage{marvosym}
\usepackage{graphicx}
\usepackage{url}
\usepackage{subfigure}
\usepackage{fancyhdr}

\DeclareMathAlphabet{\mathwee}{OT1}{cmss}{m}{sl}

\newcommand{\unit}[1]{\ensuremath{\,\mathrm{#1}}}

\newcommand{\expf}[1]{\ensuremath{\expE^{#1}}}
\renewcommand{\expf}[1]{\ensuremath{\exp\left(#1\right)}}

\newcommand{\degree}{\ensuremath{^\circ}}

\newcommand{\Om}{\ensuremath{\Omega_\mathrm{m}}}

\newcommand{\pmdepth}{\ensuremath{\delta\varphi\,}}

\begin{document}

\title{Resolved Sideband Cooling of a Micromechanical Oscillator}
\author{A.~Schliesser, R.~Rivi\`ere, G.~Anetsberger, O.~Arcizet,
T.~J.~Kippenberg} \email{tjk@mpq.mpg.de} \affiliation{Max Planck
Institut f\"ur Quantenoptik, 85748 Garching, Germany}

\begin{abstract}
Micro- and nanoscale opto-mechanical systems---based on cantilevers
\cite{Arcizet2006a, Gigan2006}, micro-cavities
\cite{Kippenberg2005,Schliesser2006} or macroscopic mirrors
\cite{Corbitt2007, Cohadon1999}---provide radiation pressure
coupling \cite{Braginsky2001} of optical and mechanical degree of
freedom and are actively pursued for their ability to explore
quantum mechanical phenomena of macroscopic objects
\cite{Schwab2005, Mancini2002}. Many of these investigations require
preparation of the mechanical system in or close to its quantum
ground state. In the past decades, remarkable progress in ground
state cooling has been achieved for trapped ions \cite{Diedrich1989,
Monroe1995} and atoms confined in optical lattices
\cite{Boozer2006,Hamann1998}, enabling the preparation of
non-classical states of motion \cite{Meekhof1996} and Schr\"odinger
cat states \cite{Monroe1996}. Imperative to this progress has been
the technique of \emph{resolved sideband cooling} \cite{Dehmelt1976,
Wineland1975, Neuhauser1978}, which allows overcoming the inherent
temperature limit of Doppler cooling \cite{Wineland1979} and
necessitates a harmonic trapping frequency which exceeds the atomic
species' transition rate. The recent advent of cavity back-action
cooling \cite{Braginsky2002} of mechanical oscillators by radiation
pressure has followed a similar path with Doppler-type cooling being
demonstrated \cite{Arcizet2006a, Gigan2006, Schliesser2006,
Corbitt2007, Karrai2006}, but lacking inherently the ability to
attain ground state cooling as recently predicted
\cite{Marquardt2007, Wilson-Rae2007}. Here we demonstrate for the
first time resolved sideband cooling of a mechanical oscillator. By
pumping the first lower sideband of an optical microcavity
\cite{Armani2003}, whose decay rate is more than twenty times
smaller than the eigen-frequency of the associated mechanical
oscillator, cooling rates above 1.5 MHz are attained, exceeding the
achievable rates in atomic species \cite{Diedrich1989}. Direct
spectroscopy of the motional sidebands reveals 40-fold suppression
of motional increasing processes, which could enable attaining final
phonon occupancies well below unity ($<0.03$). Elemental
demonstration of resolved sideband cooling as reported here, should
find widespread use in opto-mechanical cooling experiments and
represents a key step to attain ground state cooling of macroscopic
mechanical oscillators \cite{Schwab2005}. Equally important, this
regime allows realization of motion measurement with an accuracy
exceeding the standard quantum limit by two mode pumping
\cite{Braginsky1992} and could thereby allow preparation of
non-classical states of motion.
\end{abstract}

\maketitle

In atomic laser cooling, the lowest temperature which can be
attained for a trapped ion (or atom) whose harmonic trapping
frequency $\Om$ is smaller than its decay rate $\gamma$ is given by
$T_\mathrm{D}\cong \hbar \gamma/4 k_\mathrm{B}$, the \emph{Doppler
limit} \cite{Wineland1979}. In this "weak binding" regime
\cite{Wineland1979} (cf. Fig 1a), the minimum average occupation
number in the harmonic trapping potential is %
$n_\mathrm{min} \approx \gamma/4 \Om\gg 1$, which implies that the
atoms' harmonic motion cannot be cooled to the quantum ground state.
On the other hand, much lower occupation can be attained in the
\emph{resolved sideband limit}. Resolved sideband cooling
\cite{Dehmelt1976, Wineland1975} is possible when a harmonically
bound dipole such as an atom or ion exhibits a trapping frequency
$\Om\gg\gamma$, thereby satisfying the so called "strong binding
condition" \cite{Wineland1979}. The physics behind resolved sideband
cooling can be understood as follows. Owing to its harmonic motion,
a spatially oscillating excited atom will emit phase modulated
radiation $E(t)=E\, \cos\left(\omega t+\beta\cos(\Om t)\right)$ with
a modulation depth given by $\beta=k \cdot x$, where
$k=2\pi/\lambda$, $\lambda$ is the wavelength in vacuum and $x$ the
displacement amplitude along the emission direction. Consequently
the emission spectrum consists of symmetric sidebands of frequencies
$\omega_0-j \Om$ (where $j=\pm 1, \pm2,\ldots$), whose intensities
are weighted by Bessel functions $|J_j(\beta)|^2$. Inversely, the
absorption spectrum as probed by an observer in the laboratory frame
will consist of a series of absorption lines, broadened due to the
finite upper state lifetime. Cooling can be achieved by tuning the
incident radiation to one of the energetically lower lying sidebands
($\omega_\mathrm{L}=\omega_0-j\Om$, $j=1,2,\ldots$, cf.\ Fig. 1a).
This entails that the atom absorbs photons of energy $\hbar
\omega_\mathrm{L}$ while it emits on average energy $\hbar\omega_0$,
thereby reducing the ion's or atom's translational energy, leading
to cooling. The lowest average occupancy that can be attained is
given by \cite{Neuhauser1978} $n_\mathrm{min}\approx\gamma^2/16
\Om^2\ll 1$, implying that the particle can be found in the ground
state most of the time, in contrast to the "weak binding" case. This
powerful cooling technique, first proposed by Dehmelt and Wineland
\cite{Wineland1975} in 1975, has been called "cooling by motional
sideband excitation" \cite{Dehmelt1976} or "sideband cooling" and
has lead directly to the remarkable demonstration of ground state
cooling of trapped ions \cite{Diedrich1989, Monroe1995} and later of
atoms in optical lattices \cite{Boozer2006, Hamann1998}. Several
interesting extensions of this method exist \cite{Leibfried2003},
including proposed techniques in which the ion (or atom) is coupled
to a mechanical system \cite{Tian2004, Wilson-Rae2004}.

\begin{figure}[h!]
\includegraphics[width=.95\linewidth]{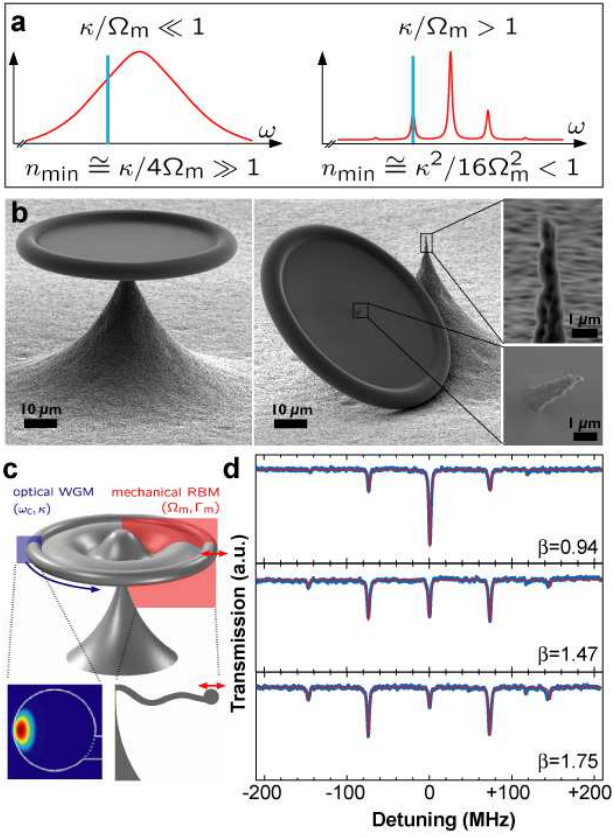}
 \caption{Resolved sideband regime of an opto-mechanical system.
   \textbf{a}~Comparison of the ``weak binding'' regime, where the
    mechanical oscillator frequency $\Om$  is smaller than the cavity decay
     rate  $\kappa$, and the ``strong binding'' regime with  $\Om\gg\kappa$. The minimum
     achievable average phonon occupancy of the mechanical oscillator is
     denoted by $n_\mathrm{min}$. Note that only in the limit of well resolved sidebands,
      ground state cooling ($n_\mathrm{min}\ll 1$) is possible. \textbf{b}~Scanning electron micrograph
      of the opto-mechanical system used in the present study, consisting of an ultra-high finesse
       toroidal optical micro-cavity with high-Q radial breathing modes ($Q=30,000$)
       supported by a ``needle'' pillar.  Also shown is an
       image of an \emph{intentionally} broken cavity structure, revealing the ultra
       thin silicon support pillar with a diameter of 500~nm, which
        reduces the coupling to the pillar and thus enables high mechanical
         $Q$ factors.  \textbf{c}~Schematic of the radiation pressure coupling
         of optical and mechanical mode in the toroidal microcavity. \textbf{d}~Cavity
          transmission spectrum of a microtoroid with a mechanical
          degree of freedom driven with a coherent drive (via radiation
          pressure back-action amplification \cite{Kippenberg2005} using an auxiliary laser beam,
           cf.\ SI) for different driving amplitudes.  The optical cavity
            decay rate corresponds to $\kappa/2 \pi =3.2 \unit{MHz}$ while the mechanical
             breathing mode exhibits a frequency of  $\Om/2\pi=73.5 \unit{MHz}$, thereby
             placing the system deeply into the resolved sideband regime.
              The weights of the sidebands follow the Bessel function
expansion (solid line is a fit). }\label{f:sems}
\end{figure}

Turning to the opto-mechanical setting consisting of a cavity
(optical degree of freedom, with a decay rate $\kappa$) with a
movable boundary (mechanical degree of freedom, with frequency
$\Om$) analogous reasoning can be made. If the strong binding
condition is satisfied, i.e. $\Om\gg\kappa$, the cavity will also
exhibit---owing to its harmonic motion---an absorption spectrum
which possesses both a carrier at the cavity resonance frequency and
a series of sidebands at $\omega_0+j\Om$. Their absorption strength
is again proportional to $|J_j(\beta)|^2$, where in the
opto-mechanical case of a whispering gallery mode cavity the
modulation index is given by $\beta=\omega_0 x/\Om R$, where $R$
denotes the cavity radius and $x$ the mechanical oscillation
amplitude. This factor can be rewritten as $\beta=k x 2 \pi/ \Om
\tau_\mathrm{rt}$,  implying that compared to the atomic dipole case
the coupling is enhanced by the number of cavity round trips (of
time $\tau_\mathrm{rt}$) the photon completes during one mechanical
oscillation period. Laser cooling can be achieved by tuning the
incident radiation to the first lower sideband. Thus the optical
cavity resonance now plays the role of the ion's optical transition
in the frequency up-conversion process \cite{Wilson-Rae2007}. Recent
theoretical work has demonstrated \cite{Marquardt2007,
Wilson-Rae2007} that, analogous to the ion trapping case
\cite{Wineland1979}, final occupancies in the mechanical degree of
freedom below unity can exclusively be attained in the resolved
sideband regime. In brief and as shown in Refs \cite{Marquardt2007,
Wilson-Rae2007}, the quantum mechanical cooling limit is due to the
fact that cooling proceeds both by motional increasing and
decreasing processes. Motional decreasing (increasing) processes
constitute absorption on the lower (upper) sideband and emission on
the carrier and occur with a rate $R^\mathrm{LSB}\propto \eta^2 A^-
n$ ($R^\mathrm{USB}\propto \eta^2 A^+(n+1)$), where $n$ is the
phonon occupancy, $A^\pm\propto \left((\kappa/2)^2+(\Delta\mp
\Om)^2\right)^{-1}$ and $\eta$ an effective parameter
\cite{Wilson-Rae2007} $\eta=\omega_0 x_0/\Om R$ with
$x_0=\sqrt{\hbar/m_\mathrm{eff} \Om}$ being the zero point motion of
the mechanical oscillator mode. Detailed balance \cite{Stenholm1986}
then yields the minimum average occupancy
$n_\mathrm{min}=A^+/(A^--A^+)$ neglecting reservoir heating. In the
"weak binding" limit $\Om\ll\kappa$ this prevents ground-state
cooling as $n_\mathrm{min}\approx \kappa/4 \Om \gg 1$, recovering
the Doppler limit. On the other hand, occupancies well below unity
can be attained in the resolved sideband case $\Om\gg\kappa$
yielding $n_\mathrm{min}\approx \kappa^2/16\Om^2\ll1$. Tuning a
laser with power $P$ to the first lower sideband ($\Delta=-\Om$),
the cooling rate $\Gamma_\mathrm{c}=A^--A^+$ under these conditions
is given by \cite{Schliesser2006}:
\begin{equation}
\Gamma_\mathrm{c}\approx\frac{8 \omega_0}{\Om}n^2 F^2
\frac{P}{m_\mathrm{eff}
c^2}\frac{\tau}{\tau_\mathrm{ex}}\frac{1}{1+4\tau^2\Om^2}.
\end{equation}
Here $F$ denotes the cavity finesse, $n$ the refractive index of the
cavity material, $m_\mathrm{eff}$ the effective mass of the
mechanical mode, $c$ the speed of light \emph{in vacuo},
$\tau=\kappa^{-1}$ the total photon lifetime and $\tau_\mathrm{ex}$
denotes the lifetimes only due to output coupling. Note that when
working in this regime of a highly detuned laser, the mechanical
frequency shift, or optical spring, is negligible, and the radiation
pressure force is mainly viscous. While several groups have recently
reported radiation pressure cooling of a mechanical oscillator
\cite{Arcizet2006a, Gigan2006, Schliesser2006, Corbitt2007}, these
experiments have all fallen into the regime of "weak binding", owing
to the challenge of simultaneously combining ultra high optical
finesse and high frequency vibrational modes.  In the presently
reported experiments we have overcome this limitation by
microfabricating optimized silica toroidal whispering gallery mode
(WGM) resonators which can accommodate the required high-quality
optical and mechanical modes in one and the same device (Fig. 1b).
The parameters $\Om/2\pi=73.5\unit{MHz}$ and $F=4.4\cdot 10^5$,
corresponding to $\kappa/2\pi=3.2\unit{MHz}$, place the system
system deeply into the resolved sideband regime. The repercussions
of this regime on the cavity transmission are most strikingly
observed when using an auxiliary laser to drive the mechanical
motion (using blue detuned light \cite{Kippenberg2005}, cf. SI).
Indeed, when tuning over the driven cavity, a series of optical
resonances spaced by the mechanical frequencies can be observed,
satisfying $\Om/\kappa\approx 22$ (cf.\ Fig.\ \ref{f:sems}c) which
convincingly prove that the device satisfies the "strong binding"
condition.

\begin{figure}[h!]
\includegraphics[width=\linewidth]{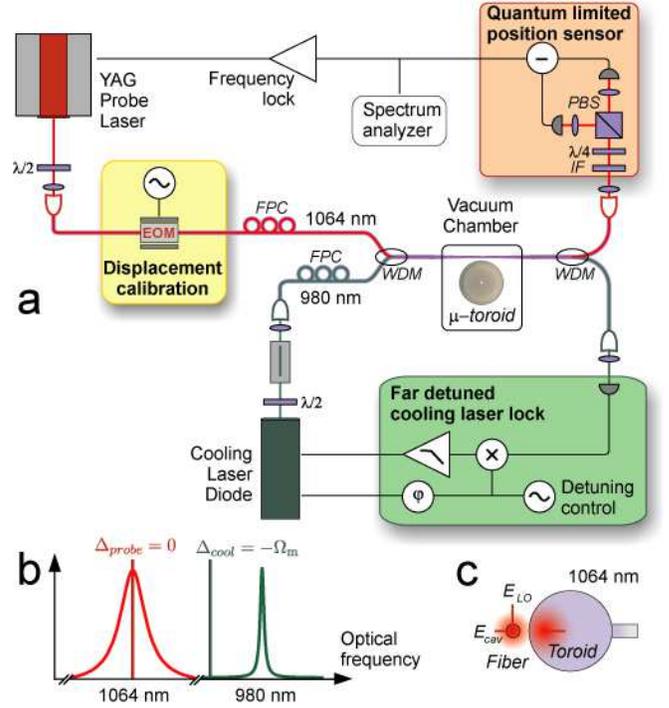}
 \caption{Schematic of the experiment. \textbf{a}~The opto-mechanical system (toroidal microcavity)
   is held in an evacuated chamber ($P<10^{-1} \unit{mbar}$), and is simultaneously
   excited by a diode laser that serves for cooling and a Nd:YAG laser
   for quantum limited monitoring of the displacement of the cavity's
    mechanical modes. \textbf{b}~By means of a frequency modulation technique,
     the cooling laser is locked far outside the cavity's WGM resonance
      to a detuning  $\Delta=-\Om$, while the monitoring laser is locked to the center
      of a different resonance using a variation of the H\"ansch-Couillaud
       technique (cf. SI). \textbf{c}~To this end, the monitoring laser is
       polarized such that only a small fraction of its power couples
       into the (polarization non-degenerate) WGM. The stronger
       orthogonal field component serves as a local oscillator in a
       polarization-sensitive detection scheme, giving rise to quantum
       noise limited displacement monitoring (cf.\ SI). The absolute
       calibration of the measured displacements is derived from a
        phase modulation of the probe laser with a known modulation depth
         (cf.\ SI). FPC: fiber polarization controller, WDM: wavelength
         division multiplexer, IF: interference filter, PBS: polarizing
          beam splitter,  $\lambda/2$, $\lambda/4$: optical retarder
          plates.}\label{f:setup}
\end{figure}

\begin{figure}[htbp]
\includegraphics[width=\linewidth]{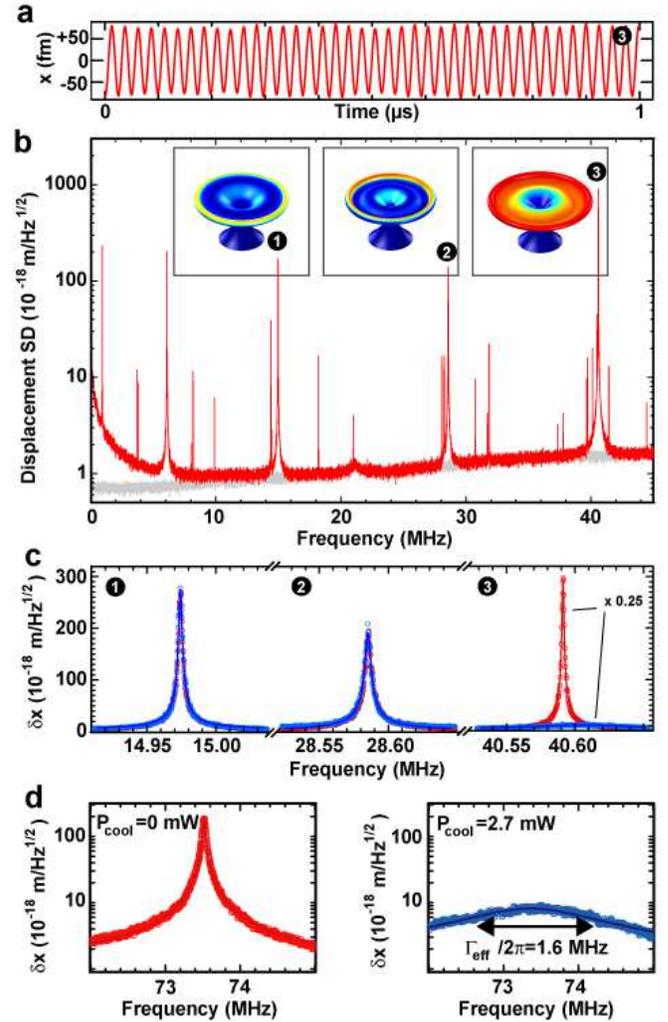}
 \caption{Resolved sideband cooling of the radial breathing mode.
   \textbf{a}~Time-domain trace of the Brownian motion of the radial
    breathing mode as observed by the monitoring laser in a 2~MHz
    spectral bandwidth around $\Om/2\pi=40.6 \unit{MHz}$.  \textbf{b}~Full
        spectrum of the displacement noise at room temperature,
        recorded with the Nd:YAG laser (red). The different peaks
        appearing in the spectrum represents the mechanical eigenmodes
         which can be identified using three-dimensional finite-element
          analysis. The modes denoted by (1,2,3) are rotationally
           symmetric mechanical modes, whose strain (colour code)
            and deformed shapes are shown in the insets. The
            background of the measurement (gray) is due to shot noise,
             its frequency dependence results from the reduced displacement
             sensitivity (for the same measured noise level) at
     frequencies exceeding the cavity's bandwidth.
     A signal-to-background ratio close to 60~dB is achieved
      at room temperature. \textbf{c}~Resolved sideband cooling with
      the cooling laser tuned to the lower sideband of the
      radially symmetric radial breathing mode (3). As evident
       only mode (3) is cooled while all other modes (of which (1)
        and (2) are shown) remain unaffected. Circles represent noise
         spectra with the cooling laser off (red) and running at
         $300 \unit{\mu W}$ (blue). Lines are Lorentzian fits.
         \textbf{d}~Cooling rates
          exceeding 1.5~MHz obtained with the 73.5-MHz RBM of a different sample.}
          \label{f:spectra}
\end{figure}

To demonstrate resolved sideband cooling, a grating-stabilized laser
diode is coupled to a high-finesse whispering-gallery mode (WGM)
near $970\unit{nm}$ of a second sample ($\Om/2\pi=40.6\unit{MHz}$,
$\kappa/2 \pi=5.8 \unit{MHz}$, $\Gamma_\mathrm{m}/2\pi=1.3
\unit{kHz}$, $m_\mathrm{eff}=10\unit{ng}$), and locked to the first
lower motional sideband (i.e. $\Delta=-\Om$) using a frequency
modulation technique and fast feedback (Fig.\ \ref{f:setup} a and
SI). The cooling caused by this laser is independently and
continuously monitored by a Nd:YAG laser ($\lambda=1064\unit{nm}$)
coupled to a different cavity mode. An adaptation of the
H\"ansch-Couillaud polarization spectroscopy technique to the
present experiment (cf. SI), allows locking the monitoring laser to
the centre of the resonance ($\Delta=0$) which, in conjunction with
low power levels, ensures that the readout laser's effect on the
mechanical oscillator motion is negligible. Simultaneously, it
provides a quantum-noise limited signal (cf.\ SI), which monitors
the displacement noise of the cavity caused by the thermal
excitation of the toroid's different mechanical modes (Fig.\
\ref{f:spectra}a,b). The sensitivity of this measurement reaches
$10^{-18}\unit{m}/\sqrt{\mathrm{Hz}}$, which is among the best
values achieved to date \cite{Arcizet2006}, and a value already
sufficient to detect the zero-point mechanical motion
$x_0=\sqrt{\hbar/m_\mathrm{eff}\Om}\sim 10^{-16}\unit{m}$ of typical
radial breathing modes (RBMs).

Increasing the cooling laser power, a clear reduction of the
displacement fluctuations is observed, characteristic of cooling.
Note that our prior work has already demonstrated unambiguously that
cooling in toroidal microcavities is \emph{solely} due to radiation
pressure \cite{Kippenberg2005, Schliesser2006} and thermal
contributions \cite{Hohberger2004} play a negligible role. We thus
observe, for the first time, resolved-sideband cooling of a
micromechanical oscillator. Simultaneous measurements on other
radially symmetric modes at lower frequencies reveal that these
remain completely unaffected (Fig.\ \ref{f:spectra}c). This
selectivity is specific to the regime of resolved sideband cooling,
in contrast to the "weak binding" case, where the  $\kappa$-wide
absorption sidebands of different mechanical modes overlap. As shown
in Fig.\ \ref{f:spectra}, the highest attained cooling rate
$\Gamma_\mathrm{c}/2\pi=1.56\unit{MHz}$ was achieved in the first
sample ($\Om/2\pi=73.5\unit{MHz}$, $\kappa/2\pi=3.2 \unit{MHz}$).
Note that of the 3~mW of launched power in the fibre only a fraction
$\sim (4\tau^2\Om^2+1)^{-1}\approx 5\cdot10^{-4}$ is coupled into
the cavity (i~e.~$1.5\unit{\mu W}$) due to the highly detuned laser.
Combining such high cooling rates with the lowest achieved reservoir
heating rates of $\Gamma_\mathrm{m}=1.3\unit{kHz}$ it appears
feasible to achieve a reduction of final ($n_\mathrm{f}$) to initial
($n_\mathrm{R}$) occupancy of
$n_\mathrm{R}/n_\mathrm{f}\cong(\Gamma_\mathrm{c}+\Gamma_\mathrm{m})/\Gamma_\mathrm{m}>10^3$.
With the demonstrated $16\Om^2/\kappa^2=7700$, this value would be
sufficient to reach $n_\mathrm{f}<0.5$ when starting at a cryogenic
temperature of 1.8~K, while still satisfying \cite{Wilson-Rae2007,
Marquardt2007} $n_\mathrm{R}/n_\mathrm{f}<Q_\mathrm{m}$ and
$\Gamma_\mathrm{c}<\kappa$. Starting from room temperature this
would lead to $n_\mathrm{f}<100$. In the actual experiment, analysis
of the \emph{integrated} calibrated displacement noise spectra via
the relation $n_\mathrm{f} \hbar \Om=\int m_\mathrm{eff}\Omega
S_x(\Omega)\, d\Omega$ indicates however significantly higher
$n_\mathrm{f}$. This discrepancy is attributed to heating by excess
phase noise from the cooling laser, which was measured to be
$\sqrt{S_\varphi}\approx 4 \unit{\mu rad}/\sqrt{\unit{Hz}}$ at radio
frequencies close to $\Om$ (cf.~SI). The resulting radiation
pressure noise limits the achievable occupancy to
$n_\mathrm{min}=\sqrt{2 k_\mathrm{B} T m_\mathrm{eff}
\Gamma_\mathrm{m} S_\varphi} R \Om /\hbar \omega_0 \approx 5200$ for
the parameters of the second sample, with which the lowest
occupancies were achieved. Note that in this case the (classical)
correlations between the laser noise and the induced displacement
fluctuations can cause "squashing" \cite{Poggio2007} artefacts if
the diode laser were also used for mechanical readout. In contrast,
the use of the independent Nd:YAG laser provides a faithful
displacement monitor with which these induced fluctuations can be
revealed. Such an analysis yields a final occupancy of
$n_\mathrm{f}\approx 5900$, in agreement with the above estimate.

A direct consequence of the resolved sideband regime is the strong
suppression of motional increasing processes, which should lead to a
significantly weaker red-sideband in the spectrum of the light
emerging from the cavity (as analyzed in Ref \cite{Wilson-Rae2007}).
To confirm this aspect, the motional sidebands generated during the
cooling cycles were probed, similar to spectroscopy of the resonance
fluorescence of a cooled ion \cite{Raab2000}. This is achieved with
a heterodyne experiment, by beating the cooling laser with a local
oscillator (cf. Fig. 4a), derived by down-shifting part of the
cooling laser light using an acousto-optic modulator at
$\Omega_\mathrm{AOM}/2\pi=200\unit{MHz}$. The beat of the local
oscillator and the cooling laser produces a modulation at
$\Omega_\mathrm{AOM}$, while the motional sidebands signals now
appear at $\Omega_\mathrm{AOM}\pm\Om$, thereby allowing the
measurement of their individual weights. Figure 4b shows the result
of this measurement for two different laser detunings. While for
excitation on cavity line-center ($\Delta=0$) the sideband
intensities are equal (i.~e. $\eta^2 A^- n\approx \eta^2 A^+(n+1)$
since $n\gg1$ and $A^-=A^+$), detuning the laser to the lower
sideband $\Delta=\-\Om$ should lead to a strong suppression of the
red sideband beat by a factor of $A^-/A^+$. In the experiment with
the 40.6-MHz sample, the detuning is chosen such that the red
sideband is still discernible above the laser noise, corresponding
to a suppression of 16 dB, providing an independent confirmation
that cooling occurs in the resolved sideband regime. Optimizing the
laser detuning, the red emission sideband could be reduced even
further. It is important to note that the ability to measure the
individual sidebands separately as demonstrated here is important
for future experiments which venture in the quantum regime. As
theoretically predicted \cite{Marquardt2007, Wilson-Rae2007}---and
in analogy to trapped ions \cite{Diedrich1989}---the weights of the
sidebands allow inferring the average motional occupation number
\cite{Wilson-Rae2007} for low occupancies by measuring the ratio of
the red and blue sidebands.

\begin{figure}[htbp]
\includegraphics[width=\linewidth]{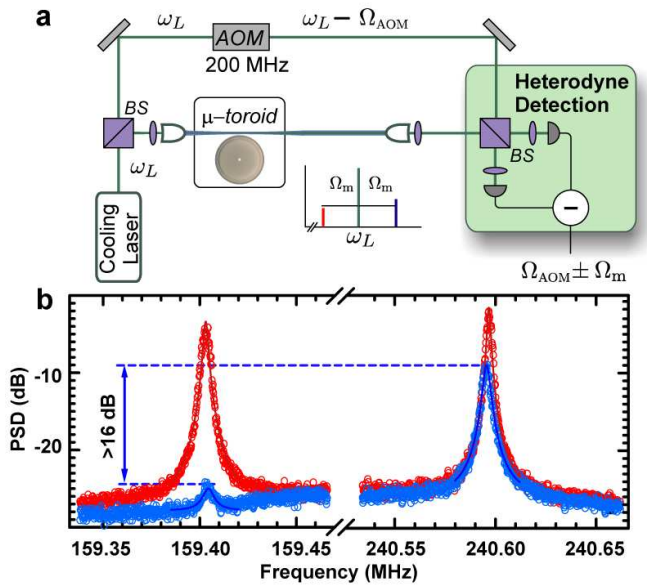}
 \caption{Motional sideband spectroscopy. \textbf{a}~Experimental setup
  used to resolve upper and lower motional sideband generated
  during interaction of the cooling laser with the cavity, similar
   to the spectroscopy of the resonance fluorescence of a laser-cooled
    ion \cite{Raab2000}. The cooling laser interacts with the optical microcavity
    whose transmission is subsequently superimposed with a second
    laser beam retrieved from the same cooling laser but down-shifted
     by 200~MHz using an acousto-optic modulator (AOM). The beating
     of the two signals is recorded using a balanced heterodyne detector,
     yielding spectral components of the lower (upper) sideband at
     $200\unit{MHz}-\Om/2\pi$ ($200 \unit{MHz}+\Om/2\pi$).
     \textbf{b}~Beat signals of the upper (anti-Stokes) and lower
       (Stokes) motional sideband, for  $\Delta=0$ (red) and a detuning close to
         $\Delta=-\Om$ (blue). The plotted electrical power spectral density (PSD) is
          proportional to the optical PSD in the sidebands.  For zero
          detuning of the pump with respect to the optical cavity the
           motional sidebands are equal in power. By tuning the laser
            to the lower sideband, which induced cooling, 40-fold
             (16 dB) reduction of the Stokes sideband is observed
              as required for achieving ground state cooling.}
\end{figure}

The regime of resolved sidebands has another important---and
counterintuitive---benefit, since the cooling rate is indeed higher
in comparison to the unresolved case. Keeping the launched power $P$
as well as $\Om$ and $R$ fixed, an increase in the cavity finesse
increases the cooling rate, until it saturates in the highly
resolved sideband case and approaches an asymptotic value (cf.~SI).
The \emph{circulating} power however continues to decrease,
mitigating undesired effects such as photo-thermal or radiation
pressure induced bi-stability or absorption induced heating. It is
important to note however, that the final occupancies for very high
optical finesse are bound by the maximum entropy flow which is given
by the cavity decay rate \cite{Marquardt2007}, i.e.
$n_\mathrm{R}/n_\mathrm{f}<\kappa/\Gamma_\mathrm{m}$.

Pertaining to the wider implications of our work, we note that
resolved sideband cooling as demonstrated here, is a key
prerequisite for ground state cooling and is an enabling step toward
observing quantum mechanical phenomena of macroscopic objects
\cite{Schwab2005, Mancini2002} such as the generation of
non-classical states of motion. The regime of resolved sidebands is
also a prerequisite to a ``continuous two transducer measurement''
scheme \cite{Braginsky1992} which allows exceeding the standard
quantum limit by $\sqrt{\Om/\kappa}$, a factor of ca.~5 for the
parameters reported here. Along these lines, continuous QND
measurements of a mechanical oscillator should produce squeezed
states of mechanical motion.

\acknowledgements The authors acknowledge discussions with T.~W.\
H\"ansch, W.\ Zwerger and I.\ Wilson-Rae. TJK acknowledges support
via an Independent Max Planck Junior Research Group Grant, a Marie
Curie Excellence Grant (JRG-UHQ), the DFG funded Nanoscience
Initiative Munich (NIM) and a Marie Curie Reintegration Grant
(RG-UHQ). The authors gratefully acknowledge J.~Kotthaus for access
to clean-room facilities for micro-fabrication.

\bibliographystyle{apsrev}
\bibliography{C:/Dokume\string~1/Albert/Eigene\string~1/Literature/microcavities}

\appendix

\begin{widetext}

\setcounter{figure}{0}%
\setcounter{equation}{0}%
\setcounter{section}{0}
\renewcommand \theequation {S\arabic{equation}}%
\renewcommand \thefigure {S\arabic{figure}}
\renewcommand \thesection {S \arabic{section}}

\newpage
\section{Sample preparation}

Silica microtoroids are fabricated from thermally grown oxide on a
silicon chip by a combination of wet- and dry-etching with a CO$_2$
reflow technique as detailed in prior work \cite{Armani2003}. Upon
completion of these initial steps, the silicon pillar supporting the
toroid is re-etched until sub-micron tip diameters are reached in
order to decouple the toroid's mechanical mode from the substrate,
yielding quality factors of 30,000 for the radial breathing mode.
Under appropriate conditions, no degradation of the optical modes is
observed in this step. Since the Nd:YAG laser has only a limited
tuning range, the toroid is subsequently subjected to repeated
low-dose radiation from a CO$_2$ laser to evaporate minute layers of
toroid material, until a suited WGM is tuned into the range
accessible with the Nd:YAG laser. The diode laser is then tuned to a
WGM with an ultra high finesse ($> 10^5$) and negligible mode
splitting.

\section{Locking of the cooling laser}
To achieve resolved sideband cooling, the cooling laser, a
grating-stabilized diode laser, has to be locked far-detuned from
the WGM resonance. To this end, it is frequency-modulated by a
applying a small rf modulation of frequency  $\Omega_\mathrm{mod}$
to the current across the diode. Similar to the Pound-Drever-Hall
(PDH) method, the resulting amplitude modulation of the light
transmitted by the cavity is demodulated with a  phase chosen to
obtain an absorptive error signal \cite{Bjorklund1983}. This creates
a steep error signal at detunings corresponding to a mechanical
mode's eigenfrequency  $\Om$, while not applying a phase modulation
resonant with the mechanical mode. Typically, a modulation frequency
of $\Omega_\mathrm{mod}=\Om-\kappa/2$  is chosen, such that
effectively a modulation sideband of the laser is locked to the red
wing of the WGM resonance, while the laser carrier pumps the
motional sideband at  $\Delta=-\Om$. To lock the laser, the error
signal is offset to choose the appropriate detuning, and fed to slow
and fast feedback branches, respectively actuating the grating and
the pump current in the diode.

\section{Cooling rate}

For the ring topology of the considered optomechanical system, the
cooling rate in the regime $\Gamma_\mathrm{c}<\kappa$ can generally
be written as \cite{Schliesser2006}
\begin{equation}
  \Gamma_\mathrm{c}=\frac{8 F^2 n^2 \omega_0}{m_\mathrm{eff} c^2
  \Om}\frac{1/\kappa \tau_\mathrm{ex}}{4\Delta^2/\kappa^2+1}
  \left(\frac{1}{4 (\Delta+\Om)^2/\kappa^2+1}-\frac{1}{4
  (\Delta-\Om)^2/\kappa^2+1}\right)P.
\end{equation}
 Here $ F$ is finesse, $n$
the refractive index, $\omega_0$ optical resonance frequency,
$m_\mathrm{eff}$ the effective mass, $c$ vacuum light speed, $\Om$
mechanical resonance frequency, $\kappa$ cavity linewidth and
therefore inverse total photon lifetime, $\tau_\mathrm{ex}$ photon
lifetime due to coupling, $\Delta$ detuning  and $P$ launched power.
All frequencies are in angular frequency units. Using the finesse of
a circular resonator $F=c/n R \kappa$ and a fixed coupling parameter
$K\equiv 1/(\kappa\tau_\mathrm{ex}-1)$, this can be re-written as
\begin{equation}
  \Gamma_\mathrm{c}=\frac{\omega_0}{\Om}\frac{1}{m_\mathrm{eff}
  R^2}\frac{K}{2(K+1)}
  \frac{\kappa^2}{\Delta^2+(\kappa/2)^2}
  \left(\frac{1}{(\Delta+\Om)^2+(\kappa/2)^2}-\frac{1}{
  (\Delta-\Om)^2+(\kappa/2)^2}\right)P.
\end{equation}
Keeping all other parameters other than $\kappa$ and $\Delta$ fixed,
it is easy to show that this expressions reaches a maximum for
$\Delta=-\Om$ and $\kappa\rightarrow 0$,
\begin{equation}
  \Gamma_\mathrm{c}^\mathrm{max}=\frac{\omega_0}{\Om^3}\frac{1}{m_\mathrm{eff}
  R^2}\frac{2 K}{K+1}P.
\end{equation}
The resolved sideband regime is \emph{per se} most ideally suited
for backaction cooling (Fig.~\ref{f:cr}), provided the linewidth is
reduced without changing the cavity radius (or length), i.e. by
increasing finesse.
\begin{figure}[h!]
\includegraphics[width=.95\linewidth]{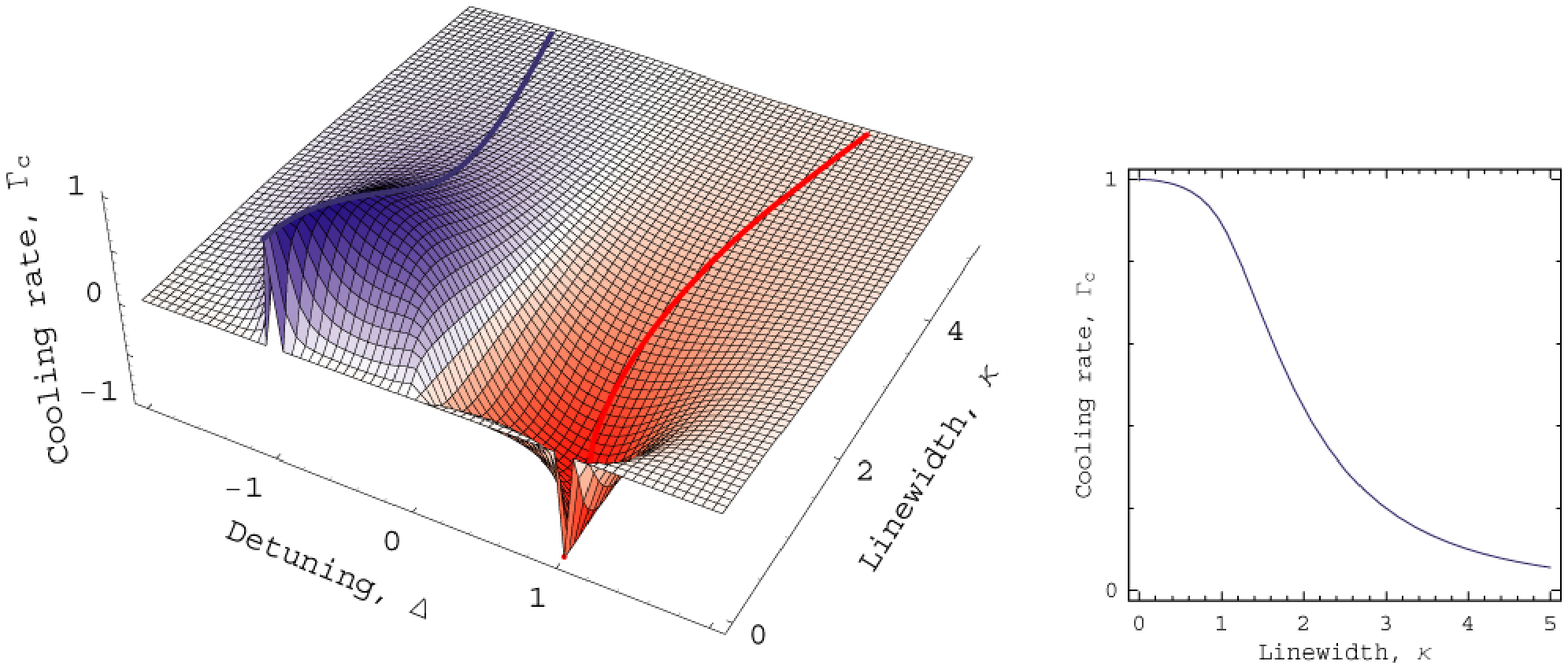}
 \caption{Left: Cooling rate $\Gamma_\mathrm{c}$, normalized to $\Gamma_\mathrm{c}^\mathrm{max}$, as a function of detuning
 $\Delta$ and cavity linewidth $\kappa$, where a mechanical resonance frequency $\Om\equiv 1$ was assumed.
 Thick blue and red lines trace the detunings which provide maximum (blue) or minimum (red) cooling rates
 for a given linewidth. Right: Normalized cooling rate achievable for a given linewidth $\kappa$ (with $\Om\equiv1$)
 when the optimum detuning is chosen.}\label{f:cr}
\end{figure}

\section{Transmission of an oscillating cavity}

Assume the cavity is oscillating such that its radius is
sinusoidally displaced by
\begin{equation}
  x(t)=x_0 \sin(\Om t).
\end{equation}
In the framework of coupled-mode theory \cite{Haus1984}, its
intracavity mode amplitude $a$ is described by
\begin{equation}
  \label{e:ode}
  \dot a(t) = \left( -\frac{\kappa}{2} + i \omega_0
  \left(1-\frac{x(t)}{R}\right) \right)
  a(t)+\frac{s}{\sqrt{\tau_\mathrm{ex}}}  e^{i \omega t}
\end{equation}
where $|a|^2$ is normalized to the intracavity energy, $|s|^2$
normalized to the incident power, $\omega_0$ the cavity resonance
frequency, $\omega$ the driving laser frequency, $R$ the mean radius
of the cavity, $\tau_\mathrm{ex}$ the lifetime due to coupling and
$\kappa$ the FWHM linewidth of the cavity in angular frequencies and
equal to the inverse total photon lifetime.

The homogeneous solution of the ordinary differential equation
(\ref{e:ode}) reads
\begin{align}
  a_\mathrm{h}(t) 
  &=A_0 \expf{  \left( -\frac{\kappa}{2} + i \omega_0\right)t+i \frac{x_0}{R}\frac{\omega_0}{\Om} \cos(\Om
  t)}
\end{align}
with a boundary-condition dependent $A_0$. One particular solution
may have the form $a_\mathrm{p}(t)=A(t) a_\mathrm{h}(t)$, satisfying
\begin{equation}
  \label{e:ODEpart}
  \dot A(t) = \frac{s}{\sqrt{\tau_\mathrm{ex}}} \expf{ \left( \frac{\kappa}{2} - i \omega_0\right)t-i \frac{x_0}{R}\frac{\omega_0}{\Om} \cos(\Om
  t) + i \omega t}.
\end{equation}
Introducing the modulation index
\begin{equation}
   \label{e:betadef}
   \beta\equiv\frac{x_0}{R}\frac{\omega_0}{\Om}
\end{equation}
the expansion of the cosine argument into Bessel functions
$
   \expf{-i \beta \cos(\Om
t)}= \sum_{n=-\infty}^{+\infty} (-i)^n J_n(\beta)\expf{i n \Om t} $
allows straightforward integration of (\ref{e:ODEpart}), yielding,
with $\Delta\equiv \omega-\omega_0$,
\begin{align}
\label{e:icfield}
  a_\mathrm{p}(t) &= \frac{s}{\sqrt{\tau_\mathrm{ex}}}
\sum_{n=-\infty}^{+\infty}\frac{(-i)^n
J_n(\beta)}{\kappa/2+i(\Delta+n \Om)}
\expf{i(\omega+n\Om)t+i\beta\cos(\Om t)}.
\end{align}

The general solution $a(t)=a_\mathrm{h}(t)+a_\mathrm{p}(t)$ of
eq.~(\ref{e:ode}) converges towards $ a_\mathrm{p}(t)$ on a
timescale of $2/\kappa$ for all $A_0$, since $ a_\mathrm{h}(t)
\propto \expf{-\kappa t/2}$, so that the steady state solution is
given by $a_p$. The field transmitted through the taper past the
cavity is given by $s_\mathrm{out}=s e^{i\omega
t}-a_\mathrm{p}/\sqrt{\tau_\mathrm{ex}} $ and the transmitted power
is
\begin{equation}
\left|s_\mathrm{out}\right|^2=\left|s e^{i\omega t}
-\frac{a_\mathrm{p}}{\sqrt{\tau_\mathrm{ex}}}\right|^2=
\left|s\right|^2-2 \mathrm{Re}\left\{
 s^* e^{-i\omega t}\cdot \frac{
 a_\mathrm{p}}{\sqrt{\tau_\mathrm{ex}}}\right\}+
\frac{|a_\mathrm{p}|^2}{\tau_\mathrm{ex}}
\end{equation}
with
\begin{align}
-2 \mathrm{Re}\left\{
 s^* e^{-i\omega t}\cdot \frac{
 a_\mathrm{p}}{\sqrt{\tau_\mathrm{ex}}}\right\}
\label{e:term2}%
&=-2 \frac{|s|^2}{\tau_\mathrm{ex}}\mathrm{Re}\left\{
\sum_{n,m}\frac{i^{m-n} J_n(\beta) J_m(\beta) \expf{i (n+m) \Om
t}}{\kappa/2+i(\Delta+n \Om)} \right\}%
\intertext{and}
\frac{|a_\mathrm{p}|^2}{\tau_\mathrm{ex}}
\label{e:term3}%
 &=\frac{|s|^2}{\tau_\mathrm{ex}^2} \left| \sum_{n,m}
\frac{i^{n-m} J_n(\beta)J_m(\beta) e^{i (n-m)\Om t} }%
{(\kappa/2+i(\Delta+n \Om))(\kappa/2-i(\Delta+m \Om))}%
 \right|^2.
\end{align}
If only the DC signal is detected, only the terms with $m=-n$  in
(\ref{e:term2}) and $m=n$ in (\ref{e:term3}) have to be considered.
Then
\begin{align}
\left|s_\mathrm{out}^\mathrm{DC}\right|^2%
&=|s|^2%
\left(%
1 %
-2 \frac{1}{\tau_\mathrm{ex}}\mathrm{Re}\left\{ \sum_{n}\frac{
J_n(\beta)^2 }{\kappa/2+i(\Delta+n \Om)} \right\}%
 +
\frac{1}{\tau_\mathrm{ex}^2} \sum_{n}
\frac{ J_n(\beta)^2}%
{(\kappa/2)^2+(\Delta+n \Om)^2}%
\right)=\\
&=|s|^2\left(1 -\frac{K\kappa^2 }{(1+K)^2}\sum_{n}\frac{
J_n(\beta)^2 }{(\kappa/2)^2+(\Delta+n \Om)^2}%
\right).
\end{align}
Therefore, the DC signal consists of a series of Lorentzian dips at
frequencies $\omega_0+n \Om$, the width of which is given by the
normal cavity linewidth $\kappa$. The depth of the dips reflects the
coupling conditions, while the modulation index $\beta=x_0
\omega_0/R \Om$ determines the relative weights of the sidebands.

In the experiment, both the diode and the Nd:YAG laser are coupled
simultaneously to the cavity, as described in Fig.~2 of the
manuscript. The Nd:YAG laser is however locked to the blue wing of
the WGM resonance, and a power sufficient to exceed the threshold
for the parametric oscillation instability is sent to the cavity.
The diode laser is then scanned over several hundred megahertz in
the frequency region of another WGM resonance. To enhance the
signal-to-noise ratio also for very low powers (at which thermal
nonlinearities \cite{Carmon2004a} are sufficiently well suppressed),
a lock-in technique is employed. The side-dips disappeared when the
Nd:YAG laser was switched off and their relative weights changed
with the detuning of this laser. From the fitted modulation depths
of $\beta=(0.94,1.47,1.75)$ oscillation amplitudes of $x_0=(5.4,
8.4, 10.0)$~pm can be inferred. For comparison, a displacement of
$25\,\mathrm{pm}$ would shift the cavity resonance frequency by one
linewidth.

\section{Displacement readout}
\subsection{Implementation}

Adapting the H\"ansch-Couillaud polarization spectroscopy technique
\cite{Hansch1980}, a dispersive error signal is created by probing
the ellipticity in polarization which is introduced to the light
propagating in the taper, when part of the light ($\vec
E_\mathrm{cav}$) couples into one of the (polarization
non-degenerate) WGM of the microtoroid. The polarization component
which does not interact with the cavity ($\vec E_\mathrm{LO}$) thus
essentially serves as a reference for the phase shift introduced by
the cavity. Sufficiently strong, it can be used as a local
oscillator, boosting all optical noises above the detector noise.

The polarization analyzer consists only of a $\lambda/4$-retarder
plate, rotated to an angle of $45\degree$ with respect to the
subsequent polarizing beam splitter (PBS). The two detected signals
represent a decomposition of the elliptically polarized light into a
basis of left- and right-handed circular polarizations. The
difference of these two signals therefore indicates the handedness
of the incoming polarization state, which indicates the respective
detuning of the laser from the WGM's line center. In addition to
this scheme \cite{Hansch1980}, in the experiment described here,
polarization mode dispersion (PMD) in the optical fibre has to be
compensated by introducing an additional pair of $\lambda/4$- and
$\lambda/2$-retarder plates before the polarization analyzer (not
shown in schematic Fig.\ 2 of the manuscript).

The complex transfer function of the taper for the cavity
polarization component is given by \cite{Haus1984}
\begin{equation}
  t(\Delta )\cdot E_\mathrm{cav}=\frac{\tau_\mathrm{ex}-\tau_0+2 i \Delta
  \tau_0\tau_\mathrm{ex}}{\tau_\mathrm{ex}+\tau_0+2 i \Delta
  \tau_0 \tau_\mathrm{ex}}\cdot E_\mathrm{cav}
\end{equation}
where $\tau_0$ is the intrinsic photon lifetime and
$\tau_\mathrm{ex}$ the lifetime due to coupling with $\kappa\equiv
\tau_0^{-1} + \tau_\mathrm{ex}^{-1}$. Using Jones matrices
$C_\Delta$, $Q$, $H$, $R_\theta$, $P$ for the effect of the cavity,
quarter- and half-waveplate, a basis rotation and polarization mode
dispersion, the detected fields $l$, $r$ on the two photodiodes are
\begin{equation}
\begin{pmatrix} l \\ r \end{pmatrix}=R_{\theta_4} (R_{\theta_3}^{-1} Q R_{\theta_3})
( R_{\theta_2}^{-1}H R_{\theta_2}) (  R_{\theta_1}^{-1} Q
R_{\theta_1} ) P  C_{\Delta} \cdot
\begin{pmatrix} E_\mathrm{cav} \\ E_\mathrm{LO} \end{pmatrix}.
\end{equation}
Adjusting the angles $\theta_i$, $P$ can be compensated such that
effectively
\begin{equation}
\begin{pmatrix} l \\ r \end{pmatrix}=R_{45\degree} Q C_{\Delta} \cdot
\begin{pmatrix} E_\mathrm{cav} \\ E_\mathrm{LO} \end{pmatrix}=%
\frac{1}{\sqrt{2}}
\begin{pmatrix}
            1& 1\\
            -1 & 1
            \end{pmatrix}
            \begin{pmatrix}
            1 & 0\\
            0 & i
            \end{pmatrix}
 \begin{pmatrix}
            t(\Delta) & 0\\
            0 & 1
            \end{pmatrix}
            \begin{pmatrix} E_\mathrm{cav} \\ E_\mathrm{LO} \end{pmatrix}=%
\frac{1}{\sqrt{2}}\begin{pmatrix}
            t(\Delta)E_\mathrm{cav}+ i E_\mathrm{LO}\\
            -t(\Delta) E_\mathrm{cav}+i E_\mathrm{LO}
            \end{pmatrix}
\end{equation}

\subsection{Sensitivity}

In the case of displacements slow enough so that intracavity power
can adiabatically follow, the H\"ansch-Couillaud error signal
$h(\Delta)=|l|^2-|r|^2$ reads
\begin{equation}
h(\Delta)=\frac{8 \tau_0^2 \tau_\mathrm{ex}
\Delta}{\tau_\mathrm{ex}^2+2
\tau_0\tau_\mathrm{ex}+\tau_0^2(1+4\Delta^2\tau_\mathrm{ex}^2)}
\sqrt{P_\mathrm{cav}P_\mathrm{LO}}.
\end{equation}

If the cavity is probed on resonance $\omega-\omega_0\equiv0$, small
displacements $x$ from an original resonator radius $R$ will cause a
detuning of $\Delta=\omega_0 x/R$ and therefore a signal of
\begin{equation}
\left.\frac{\partial}{\partial x}h(\omega_0 x/R)\right|_{x=0}\cdot
x=\frac{ 8\omega_0}{\tau_\mathrm{ex}\kappa^2
R}\sqrt{P_\mathrm{cav}P_\mathrm{LO}}\cdot x
\end{equation}
can be expected. In the case of critical coupling
$\tau_\mathrm{ex}=2/\kappa$ and for a strong local oscillator
$P_\mathrm{LO}\gg P_\mathrm{cav}$, the local oscillator's shot noise
in the differential signal $\sqrt{2} \sqrt{\eta P_\mathrm{LO}/2
/(\hbar\omega)}$ can be used to estimate the shot-noise limited
sensitivity to
\begin{equation}
x_\mathrm{min}=\frac{\lambda}{8 \pi \mathcal F \sqrt{\eta
P_\mathrm{cav}/(\hbar \omega)}},
\end{equation}
with the (ring) resonator finesse $\mathcal F= c / \kappa n R$ and
the intracavity laser wavelength $\lambda=2 \pi c/ n \omega$.
Assuming shot-noise limited operation, typical parameters $\mathcal
F=40{,}000$, $\eta=0.5$, $P_\mathrm{cav}=1\,\mathrm{\mu W}$ and
$\lambda=1064\,\mathrm{nm}/1.4$ already yield a sensitivity of
$x_\mathrm{min}=5\cdot 10^{-19} \,\mathrm{m}/\sqrt\mathrm{Hz}$.

We have confirmed experimentally that the readout laser is
shot-noise limited, and that the electronic noise in the detection
system is overwhelmed by the laser noise by more than 10 dB at all
frequencies of interest.


It is important to consider the response of the H{\"a}nsch-Couillaud
signal also at frequencies exceeding the cavity bandwidth for
measurements on high-frequency mechanical modes. Assuming small
displacement fluctuations $\beta=x_0 \omega_0 / R \Omega \ll 1$ at
Fourier frequency $\Omega$, the expression (\ref{e:icfield}) for the
intracavity field can be expanded in $\beta$ and only terms in
zeroth and first order in $\beta$ retained:
\begin{equation}
  a_\mathrm{p}(t)=a_0(t)+\mathcal{O}(\beta^1)=a_0(t)+a_1(t)+\mathcal{O}(\beta^2)=\ldots
\end{equation}
with
\begin{align}
a_0(t)&= \frac{s e^{i \omega t}}{\sqrt{\tau_\mathrm{ex}}}
\frac{J_0(\beta)^2}{\kappa/2+i\Delta}\\
a_1(t)&= \frac{s e^{i \omega t}}{\sqrt{\tau_\mathrm{ex}}} %
\left( %
  \left(%
    \frac{ -i J_1(\beta) \expf{+i \Om t}}{\kappa/2+ i
    (\Delta+\Om)} +
    \frac{ i J_{-1}(\beta) \expf{- i \Om t}}{\kappa/2+ i
    (\Delta-\Om)}
 \right)\cdot J_0(\beta)+\right.
\nonumber\\
 &\qquad+ \left. \frac{J_0(\beta)}{\kappa/2+i \Delta}\cdot
 \left(i J_1(\beta)\expf{+i \Om t}-i J_{-1}(\beta)\expf{-i \Om
 t}\right)
 \right)=\\
&=\frac{s e^{i \omega t}}{\sqrt{\tau_\mathrm{ex}}} %
\frac{J_0(\beta)J_1(\beta)\Om}{\kappa/2+i \Delta} %
\left( %
-\frac{ \expf{+i \Om t}}{ %
\kappa/2+ i(\Delta+\Om)}+%
 \frac{\expf{-i \Om t}}{ %
\kappa/2+ i(\Delta-\Om)}%
\right)
\end{align}
Using $J_0(\beta)\cong 1$ and $J_1(\beta)\cong\beta/2$ for small
$\beta$ and the definition (\ref{e:betadef}), one finds
\begin{equation}
 a(t)\cong\frac{E_\mathrm{cav}}{\sqrt{\tau_\mathrm{ex}}}%
 \left(
 \frac{1}{\kappa/2+i\Delta}-
\frac{x}{R}\frac{\omega_0}{2}\frac{1}{\kappa/2+i \Delta} %
\left( %
\frac{ \expf{+i \Omega t}}{ %
\kappa/2+ i(\Delta+\Omega)}-%
 \frac{\expf{-i \Omega t}}{ %
\kappa/2+ i(\Delta-\Omega)}%
\right) \right).
\end{equation}
The detected fields behind the polarization analyzer are
\begin{equation}
\begin{pmatrix} l \\ r \end{pmatrix}=R_{45^\circ} \cdot Q \cdot
\begin{pmatrix} -a /\sqrt{\tau_\mathrm{ex}} +E_\mathrm{cav} \\
E_\mathrm{LO}\end{pmatrix}= \frac{1}{\sqrt{2}}
\begin{pmatrix}
-a /\sqrt{\tau_\mathrm{ex}} +E_\mathrm{cav}+ i E_\mathrm{LO}\\
a /\sqrt{\tau_\mathrm{ex}} -E_\mathrm{cav}+ i E_\mathrm{LO}
\end{pmatrix}.
\end{equation}
If the readout laser is locked to line center, $\omega=\omega_0$,
the signal on one photodiode reads
\begin{align}
  |l|^2&=\frac{1}{2}\left|
  -\frac{E_\mathrm{cav}}{\tau_\mathrm{ex}}\frac{2}{\kappa}
\left( 1-\frac{x_0}{R}\frac{\omega_0}{2}\left( \frac{e^{+i\Omega
t}}{\kappa/2+i \Omega}-\frac{e^{-i\Omega t}}{\kappa/2-i
\Omega}\right) \right) + E_\mathrm{cav}+i E_\mathrm{LO}
  \right|^2,
    \intertext{for a strong local oscillator
  $|E_\mathrm{LO}|\gg |E_\mathrm{cav}|$ (and rigorously for critical
  coupling $\tau_\mathrm{ex}=2/\kappa$) simplifying to}
  &\cong|l|^2_\mathrm{DC}-\frac{1}{2}\cdot 2 \mathrm{Re}\left\{
  i\frac{  E_\mathrm{LO} E_\mathrm{cav}}{\tau_\mathrm{ex}} \frac{2}{\kappa}
  \frac{x_0}{R}\frac{\omega_0}{2}\left( \frac{e^{+i\Omega
t}}{\kappa/2+i \Omega}-\frac{e^{-i\Omega t}}{\kappa/2-i
\Omega}\right)  \right\}=\\
  &=|l|^2_\mathrm{DC}+\frac{\sqrt{P_\mathrm{LO}P_\mathrm{cav}}}{\tau_\mathrm{ex}}
 \frac{2}{\kappa}
  \frac{x_0}{R}\omega_0
   \frac{(\kappa/2)\sin(\Omega t) -\Omega\cos(\Omega t)
  }{(\kappa/2)^2+\Omega^2}
\end{align}
An analogous calculation for $r$ yields
\begin{equation}
\label{e:hcdyn} h(t)=2
\frac{\sqrt{P_\mathrm{LO}P_\mathrm{cav}}}{\tau_\mathrm{ex}}
 \frac{2}{\kappa}
  \frac{x_0}{R}\omega_0
   \frac{(\kappa/2)\sin(\Omega t) -\Omega\cos(\Omega t)
  }{(\kappa/2)^2+\Omega^2},
\end{equation}
reproducing the adiabatic response in the corresponding limit
$\Omega\ll\kappa/2$. This modulated optical power generates a
modulated (differential) photocurrent $I=e \eta h/(\hbar\omega)$.
After amplification by a transimpedance amplifier of gain $g [V/A]$,
the \emph{power} $P_\mathrm{rf}=|g I|^2/R_\mathrm{term}$ (with
$R_\mathrm{term}$ the termination resistance) of the radio-frequency
signal is  spectrally analyzed. Then the sine and cosine terms add
in quadrature, and
\begin{equation}
P_\mathrm{rf}(\Omega)\propto \frac{(\kappa/2)^2+\Omega^2}
{((\kappa/2)^2+\Omega^2)^2}= \frac{1} {(\kappa/2)^2+\Omega^2}.
\end{equation}
This reduced response of the readout signal has to be considered
when the measured photocurrent spectral density is converted to a
displacement spectral density. It is also the reason why the
shot-noise limited sensitivity is not flat in Fourier frequencies,
instead it scales as
\begin{equation}
x_\mathrm{min}(\Omega)=\frac{\lambda}{8 \pi \mathcal F \sqrt{\eta
P_\mathrm{cav}/(\hbar \omega)}}
\sqrt{1+\frac{\Omega^2}{(\kappa/2)^2}}.
\end{equation}

\subsection{Calibration}

To calibrate the measured displacements, the light of the readout
laser is phase-modulated using an electro-optic modulator. If the
modulation depth is $\delta\varphi$ and the modulation frequency
$\Omega$, the field sent to the cavity can approximately written as
$E_\mathrm{cav} (1+\frac{\pmdepth}{2} e^{i \Omega
t}-\frac{\pmdepth}{2} e^{-i \Omega t})$ and the local oscillator,
\emph{which also gets modulated}, as $E_\mathrm{LO}
(1+\frac{\pmdepth}{2} e^{i \Omega t}-\frac{\pmdepth}{2} e^{-i \Omega
t})$. The field detected by one photodiode then reads
\begin{align}
l&=\frac{1}{\sqrt{2}}\left(
\left(1-\frac{1}{\tau_\mathrm{ex}}\frac{1}{\kappa/2+i
\Delta}\right)E_\mathrm{cav}
+\frac{\pmdepth}{2}\left(1-\frac{1}{\tau_\mathrm{ex}}\frac{1}{\kappa/2+i
(\Delta+\Omega)}\right)E_\mathrm{cav}e^{i\Omega
t}-\right.\nonumber\\&\left.\qquad-\frac{\pmdepth}{2}\left(1-\frac{1}{\tau_\mathrm{ex}}\frac{1}{\kappa/2+i
(\Delta-\Omega)}\right)E_\mathrm{cav}e^{-i\Omega t} \right)+i
\left(1+\frac{\pmdepth}{2}e^{i \Omega t}-\frac{\pmdepth}{2}e^{-i
\Omega t}\right)E_\mathrm{LO}.
\end{align}
If the laser is locked to line center, $\Delta\equiv 0$ and
\begin{align}
|l|^2&=|l|^2_\mathrm{DC}+\frac{1}{2}\cdot 2 \mathrm{Re}%
\left\{%
\frac{\pmdepth}{2}\left(1-\frac{1}{\tau_\mathrm{ex}}\frac{1}{\kappa/2+i\Omega}\right)E_\mathrm{cav}e^{i\Omega
t}(-i E_\mathrm{LO})%
-\frac{\pmdepth}{2}\left(1-\frac{1}{\tau_\mathrm{ex}}\frac{1}{\kappa/2-i\Omega}\right)E_\mathrm{cav}
e^{-i\Omega t}(-i E_\mathrm{LO})+%
\right.\nonumber\\&\left.\qquad
+\left(1-\frac{1}{\tau_\mathrm{ex}}\frac{2}{\kappa}\right)E_\mathrm{cav}%
\left(-i\frac{\pmdepth}{2}e^{-i\Omega t}\right)%
-\left(1-\frac{1}{\tau_\mathrm{ex}}\frac{2}{\kappa}\right)E_\mathrm{cav}%
\left(i\frac{\pmdepth}{2}e^{i\Omega t}\right)%
\right\}=\\
 &=|l|^2_\mathrm{DC}+
 \frac{1}{\tau_\mathrm{ex}}\pmdepth \Omega \sqrt{P_\mathrm{cav}
 P_\mathrm{LO}}\frac{2}{\kappa}\frac{(\kappa/2)\cos(\Omega
 t)+\Omega \sin(\Omega t)}{(\kappa/2)^2+
 \Omega^2},
\end{align}
such that the H\"ansch-Couillaud signal resulting from this phase
modulation is
\begin{equation}
h(t)=2 \frac{\sqrt{P_\mathrm{LO}P_\mathrm{cav}}}{\tau_\mathrm{ex}}
\pmdepth\Omega
 \frac{2}{\kappa}
   \frac{(\kappa/2)\cos(\Omega t) +\Omega\sin(\Omega t)
  }{(\kappa/2)^2+\Omega^2}.
\end{equation}
Comparison with the signal (\ref{e:hcdyn}) arising from harmonic
displacement of amplitude $x_0$, one can infer that phase modulation
by $\pmdepth$ leads to the same signal as a harmonic displacement of
amplitude $x_0 \omega_0/R\Omega$---independent of cavity bandwidth,
coupling conditions and readout power (note again that that sin- and
cos-quadratures of the modulation add in quadrature). This can be
used to calibrate the measured spectra by injecting a phase
modulation of known depth at one particular frequency.

Experimentally, the modulation is produced using a fibre-coupled
$\mathrm{LiNbO}_3$-waveguide, to which a modulated voltage is
applied. The depth of modulation for a certain applied voltage was
derived by measuring the relative strengths of radio-frequency
sidebands of a heterodyne beat of the modulated Nd:YAG laser with an
independent diode laser at 1064~nm. A value of
$17\degree/\mathrm{V}$ was found, in good a agreement with the
specified value. We have checked that the residual amplitude
modulation created in the modulator is negligible.

\section{Laser noise heating}

While the cooling laser, an external cavity grating-stabilized
semiconductor laser, exhibits low intensity noise, its output was
found to contain significant excess frequency noise even at the
relevant Fourier frequencies above 10~MHz, as reported also by other
groups \cite{Zhang1995}. Measurements against an independent cavity
indicate frequency noise on the order of $200
\unit{Hz}/\sqrt{\unit{Hz}}$ for the Fourier frequencies of interest,
corresponding to a phase noise of about $4 \unit{\mu
rad}/\sqrt{\unit{Hz}}$. To estimate the resulting heating effect it
is first assumed the laser carries a sinusoidal phase modulation of
depth $\pmdepth$ at a Fourier frequency of $\Om$, the resonance
frequency of the mechanical oscillator. Using the Bessel expansion
for small $\pmdepth$, the incoming light field can be written as
\begin{equation}
  s_\mathrm{in}(t)\cong \left(1 +\frac{\pmdepth}{2} e^{+i \Om t}- \frac{\pmdepth}{2} e^{-i \Om t} \right) s e^{i \omega
  t}
\end{equation}
with the incoming amplitude $s$ normalized such that $|s|^2\equiv P$
is the optical power. Then the intracavity mode amplitude $a$ (with
$|a|^2$ normalized to intracavity energy) can be written as
\cite{Haus1984}
\begin{equation}
  a(t)=\left(
  \frac{1}{\kappa/2+i \Delta}+
  \frac{\pmdepth e^{+i\Om t}/2 }{\kappa/2+i
  (\Delta+\Om)}-
  \frac{\pmdepth e^{-i\Om t}/2 }{\kappa/2+i (\Delta-\Om)}
  \right)\frac{s e^{i \omega t}}{\sqrt{\tau_\mathrm{ex}}}.
\end{equation}
In the deeply resolved sideband regime $\kappa\ll\Om$ this
simplifies to
\begin{equation}
  a(t)\cong\left(
  \frac{1}{-i \Om}+ \frac{\pmdepth e^{+i\Om t}/2}{\kappa/2}
  \right) \frac{s e^{i \omega t}}{\sqrt{\tau_\mathrm{ex}}}.
\end{equation}
if a detuning $\Delta=-\Om$ is  assumed. The intracavity energy
therefore contains a modulation $\propto \sin(\Om t)$ of amplitude
$\pmdepth/\Om |s|^2$ for critical coupling
$\tau_\mathrm{ex}=2/\kappa$. This gives rise to a modulated force of
amplitude $\delta F=\pmdepth P/\Om R$, with $R$ the cavity radius.
Substituting $\pmdepth$ with the measured phase noise \emph{spectral
density}, a noise force spectral density (SD) of
\begin{equation}
 S_{F_\mathrm{noise}}\cong \frac{S_\varphi P^2}{\Om^2  R^2}
\end{equation}
is obtained. This value can be considered constant for Fourier
frequencies around $\Om$ as long as $\Gamma_\mathrm{eff}\ll\kappa$.
An analogous approach yields a random force of spectral density $S_I
P^2/\Om^2 R^2$ for a laser with a relative (i.~e. normalized to
$I^2$) intensity noise spectral density $S_I$. In the case of the
employed laser we have measured that $S_\varphi$ exceeds $S_I$ by
nearly four orders of magnitude, so that $S_I$ can be negelected.
Comparison with the thermal force SD $S_\mathrm{th}=2 k_\mathrm{B} T
m_\mathrm{eff} \Gamma$ then yields a power-dependent temperature of
the laser
\begin{equation}
  T_\mathrm{laser}=\frac{1}{2 k_\mathrm{B} m_\mathrm{eff}  \Gamma}\frac{ S_\varphi P^2}{\Om^2 R^2}
\end{equation}
Simultaneously, the laser cools the mode by modifying its
susceptibility to the different noisy forces, with a rate
$\Gamma_\mathrm{cool}\cong\omega P/\Om^3 m_\mathrm{eff}  R^2$ in the
approximations described above. For a reservoir temperature $T$, the
effective temperature of the mode is
\begin{equation}
  T_\mathrm{eff}\cong\frac{\Gamma}{\Gamma_\mathrm{cool}}
  \left(T + T_\mathrm{laser}  \right).
\end{equation}
The apparent contradiction of such a calculation with the properties
of the \emph{intensive} thermodynamical variable ``temperature'' is
resolved by realizing that the quoted temperatures are effective
temperatures representing uncorrelated random forces, which can be
added in quadrature. For the optimum laser power of
$P_\mathrm{opt}=\sqrt{2 k_\mathrm{B} T m_\mathrm{eff}  \Gamma/
S_\varphi} \cdot R\Om$, the lowest achievable phonon occupancy is
therefore found to be
\begin{equation}
  n_\mathrm{min}\cong \sqrt{2 k_\mathrm{B} T m_\mathrm{eff}  \Gamma S_\varphi}\frac{R \Om}{\hbar \omega}.
\end{equation}
With $T=300\unit{K}$, $m_\mathrm{eff} =10 \unit{ng}$,
$\Gamma/2\pi=1.3\unit{kHz}$, $\sqrt{S_\varphi}=4 \unit{\mu rad}
/\sqrt{\unit{Hz}}$, $k_\mathrm{B}=1.4 \cdot 10^{-23}
\unit{J}/\unit{K}$, $R=38 \unit{\mu m}$, $\Om/2 \pi=40.6
\unit{MHz}$, $\hbar=1.05\cdot 10^{-34} \unit{J\,s}$ and $\omega/2
\pi=300 \unit{THz}$, a minimum phonon number of about $5200$ is
found.

\end{widetext}

\end{document}